\documentclass[epj,nopacs]{svjour}
\pdfoutput=1
\usepackage{graphicx,hyperref,wrapfig,cite,color,subfigure,amssymb,amsmath,epsfig}
\usepackage[utf8]{inputenc}
\usepackage[english]{babel}
\hypersetup{
     colorlinks=true,        % false: boxed links; true: colored links
     linkcolor=red,          % color of internal links
     citecolor=green,        % color of links to bibliography
     filecolor=magenta,      % color of file links
     urlcolor=blue           % color of external links
}
\newcommand{\newc}{\newcommand}
\def\u#1{\verb!#1!\endgroup}

\newc{\HW}{\mbox{\textsf{HERWIG}}}
\newc{\Hw}{\mbox{\textsf{Herwig}}}
\newc{\KrkNLO}{\textsf{KrkNLO}}
\newc{\OL}{\textsf{OpenLoops}}
\newc{\Collier}{\textsf{Collier}}
\newc{\GoSam}{\textsf{GoSam}}
\newc{\TAUOLA}{\textsf{TAUOLA}}
\newc{\ThePEG}{\textsf{ThePEG}}
\newc{\boost}{\textsf{BOOST}}
\newc{\HepMC}{\textsf{HepMC}}
\newc{\Rivet}{\textsf{Rivet}}
\newc{\lhapdf}{\textsf{LHAPDF}}
\newc{\HWPP}{\mbox{\textsf{Herwig++}}}
\newc{\evt}{\textsf{EvtGen}}
\newc{\fortran}{\textsf{FORTRAN}}
\newc{\decayer}{\textsf{Decayer}}
\newc{\matchbox}{\textsf{Matchbox}}

\newc{\HWPPClass}[1]{\href{https://herwig.hepforge.org/doxygen/classHerwig_1_1#1.html}{\textsf{#1}}}
\newc{\ThePEGClass}[1]{\href{https://thepeg.hepforge.org/doxygen/classThePEG_1_1#1.html}{\textsf{#1}}}
\newc{\HWPPParameter}[2]{\href{https://herwig.hepforge.org/doxygen/#1Interfaces.html\##2}{{\bf #2}}}
\newc{\ThePEGParameter}[2]{\href{https://thepeg.hepforge.org/doxygen/#1Interfaces.html\##2}{{\bf #2}}}
\newc{\HWPPParameterValue}[3]{\href{https://herwig.hepforge.org/doxygen/#1Interfaces.html\##2}{{\bf [#2=#3]}}}
\newc{\HWPPParameterValueB}[3]{\href{https://herwig.hepforge.org/doxygen/#1Interfaces.html\##2}{{\bf [#3]}}}
\newc{\ThePEGParameterValue}[3]{\href{https://thepeg.hepforge.org/doxygen/#1Interfaces.html\##2}{{\bf [#2=#3]}}}

\preprint{
CERN-PH-TH-2017-109\\ % Sorted by length! That looks nicest
MAN/HEP/2017/08\\
UWTHPH-2017-10\\
IFJPAN-IV-2017-7\\
NIKHEF 2017-026 \\
HERWIG-2017-02\\
KA-TP-19-2017\\
MCnet-17-08\\
IPPP/17/40
}

\title{Herwig 7.1 Release Note}

\author{
Johannes Bellm\inst{1}\and
Stefan Gieseke\inst{2}\and
David Grellscheid\inst{1}\and
Patrick Kirchgae\ss{}er\inst{2}\and
Frash\"{e}r Loshaj\inst{2}\and
Graeme Nail\inst{3}\and
Andreas Papaefstathiou\inst{4,5}\and
Simon Pl\"atzer\inst{1,3,6}\and
Radek Podskubka\inst{2}\and
Michael Rauch\inst{2}\and
Christian Reuschle\inst{7}\and
Peter~Richardson\inst{1,8}\and
Peter Schichtel\inst{1}\and
Michael H. Seymour\inst{3}\and
Andrzej Si\'odmok\inst{9}\and
Stephen Webster\inst{1}
}
\institute{
IPPP, Department of Physics, Durham University\and
Institute for Theoretical Physics, Karlsruhe Institute of Technology\and
Particle Physics Group, School of Physics and Astronomy, University of
Manchester\and
NIKHEF, Theory Group,\and
Institute for Theoretical Physics and Delta Institute for Theoretical Physics,
University of Amsterdam\and
Particle Physics, Faculty of Physics, University of Vienna\and
HEP Theory Group, Department of Physics, Florida State University\and 
CERN, PH-TH, Geneva\and
The Henryk Niewodniczanski Institute of Nuclear Physics in Cracow, Polish
Academy of Sciences
}

\date{\today}

\abstract{A new release of the Monte Carlo event generator Herwig (version
  7.1) is now available. This version introduces a number of improvements,
  notably: multi-jet merging with the dipole shower at LO and NLO QCD; a new
  model for soft interactions and diffraction; improvements to mass effects
  and top decays in the dipole shower, as well as a new tune of the
  hadronisation parameters.}

\begin{document}\sloppy

\maketitle

\section{Introduction}

\Hw\ is a multi purpose particle physics event generator. Its latest version,
\Hw\ 7.0 \cite{Bellm:2015jjp}, which was based on a major development
of the \HWPP
\cite{Bahr:2008pv,Bahr:2008tx,Bahr:2008tf,Gieseke:2011na,
  Arnold:2012fq,Bellm:2013lba} branch, has superseded the \HWPP\ 2.x and
\HW\ 6.x versions. Building on the technology and experience gained with the
higher-order improvements provided by \Hw\ 7.0, a major follow-up release,
\Hw\ 7.1 is now available and provides multijet merging at next-to-leading
order QCD \cite{Bellm:2017xxx} as one of its main new features. In addition,
the new version includes several improvements to the soft components of the
simulation, amongst other changes and physics capabilities, which we will
highlight in this release note. Please refer to the \HWPP\ manual
\cite{Bahr:2008pv}, the \Hw\ 7.0 \cite{Bellm:2015jjp} as well as this release
note when using the new version of the program.
Studies or analyses that rely on a particular feature of the program should
also reference the paper(s) where the physics of that feature was first
described.

\subsection{Availability}

The new version, as well as older versions of the \Hw\ event generator can be
downloaded from the website
\texttt{\href{https://herwig.hepforge.org/}{https://herwig.hepforge.org/}}.
We strongly recommend using the \texttt{bootstrap} script provided for the
convenient installation of \Hw\ and all of its dependencies, which can be
obtained from the same location. On the website, tutorials and FAQ sections are
provided to help with the usage of the program. Further enquiries should be
directed to \texttt{herwig@projects.hepforge.org}.  \Hw\ is released under the
GNU General Public License (GPL) version 3 and the MCnet guidelines for the
distribution and usage of event generator software in an academic setting, see
the source code archive or
\texttt{\href{http://www.montecarlonet.org/index.php?p=Publications/Guidelines}
  {http://www.montecarlonet.org/}}.

\subsection{Prerequisites and Further Details}

\Hw\ 7.1 is built on the same backbone and dependencies as its predecessor
\Hw\ 7.0, and uses the same method of build, installation and run
environment. No major changes should hence be required in comparison to a
working \Hw\ 7.0 installation. Some of the changes, though,
might require different compiler versions, particularly our switch to the
C++~11 standard.
The tutorials at
\texttt{\href{https://herwig.hepforge.org/tutorials/}{https://herwig.hepforge.org/tutorials/}}
have been extended and adapted to the new version and serve as the primary
reference for physics setups and as a user manual until a comprehensive
replacement for the detailed manual \cite{Bahr:2008pv} is available.
  
\section{Multijet Merging}  

Based on the \matchbox\ development \cite{Platzer:2011bc} which is central to
the NLO {\it matching} capabilities of \Hw, a multijet {\it merging} algorithm
detailed in \cite{Bellm:2017xxx} has been implemented together with the dipole
shower algorithm and based on an improved, unitarised merging prescription
following the proposal set out in \cite{Platzer:2012bs}. The algorithm is able
to merge cross sections for multiple jet production at the NLO QCD level, and
has been tested with a range of standard model processes such as vector boson
or Higgs boson plus jets production, top pair production, and pure jet
production.

\begin{figure}[t]
  \centering
  \scalebox{.7}{\includegraphics{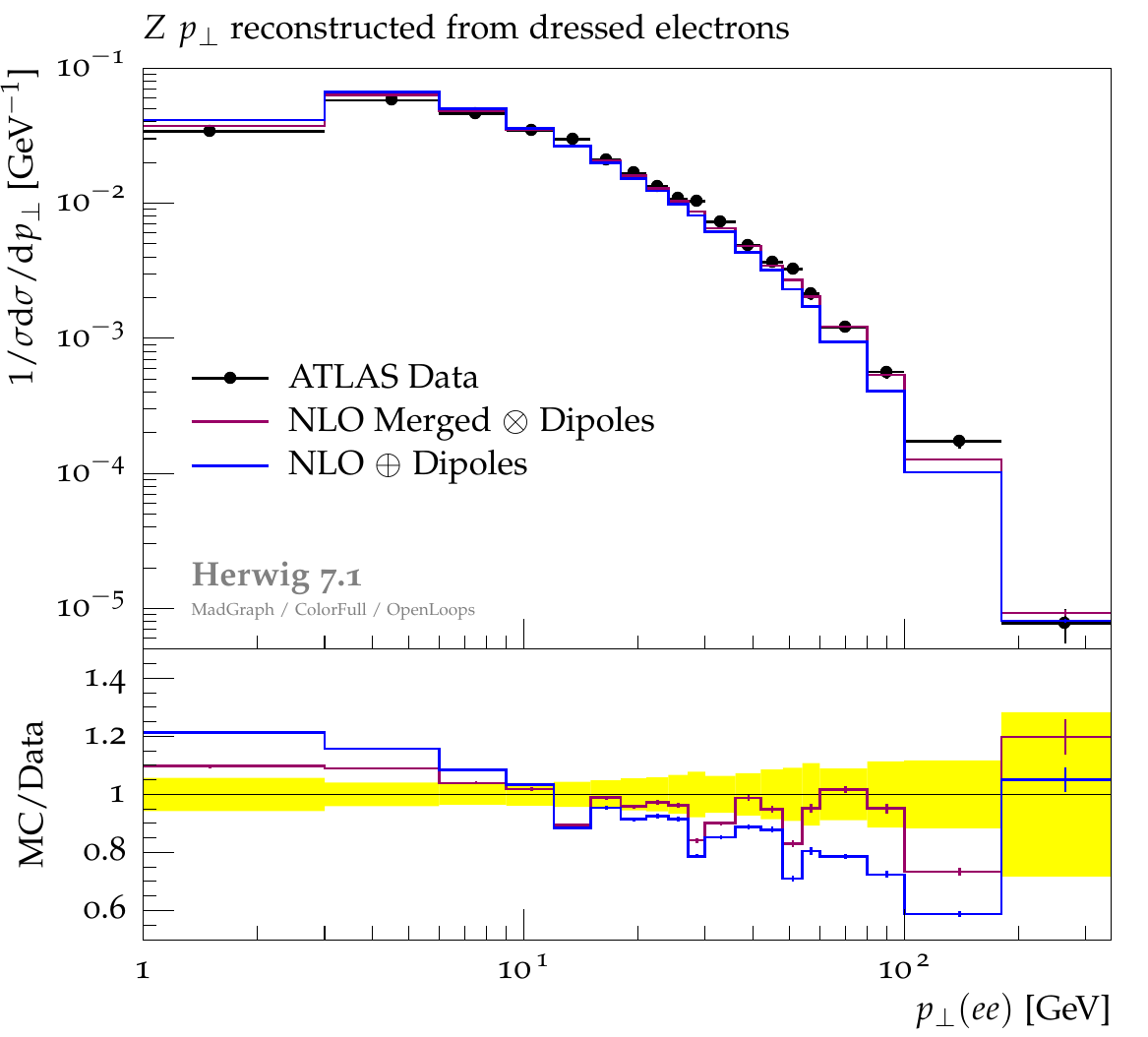}}
  \scalebox{.7}{\includegraphics{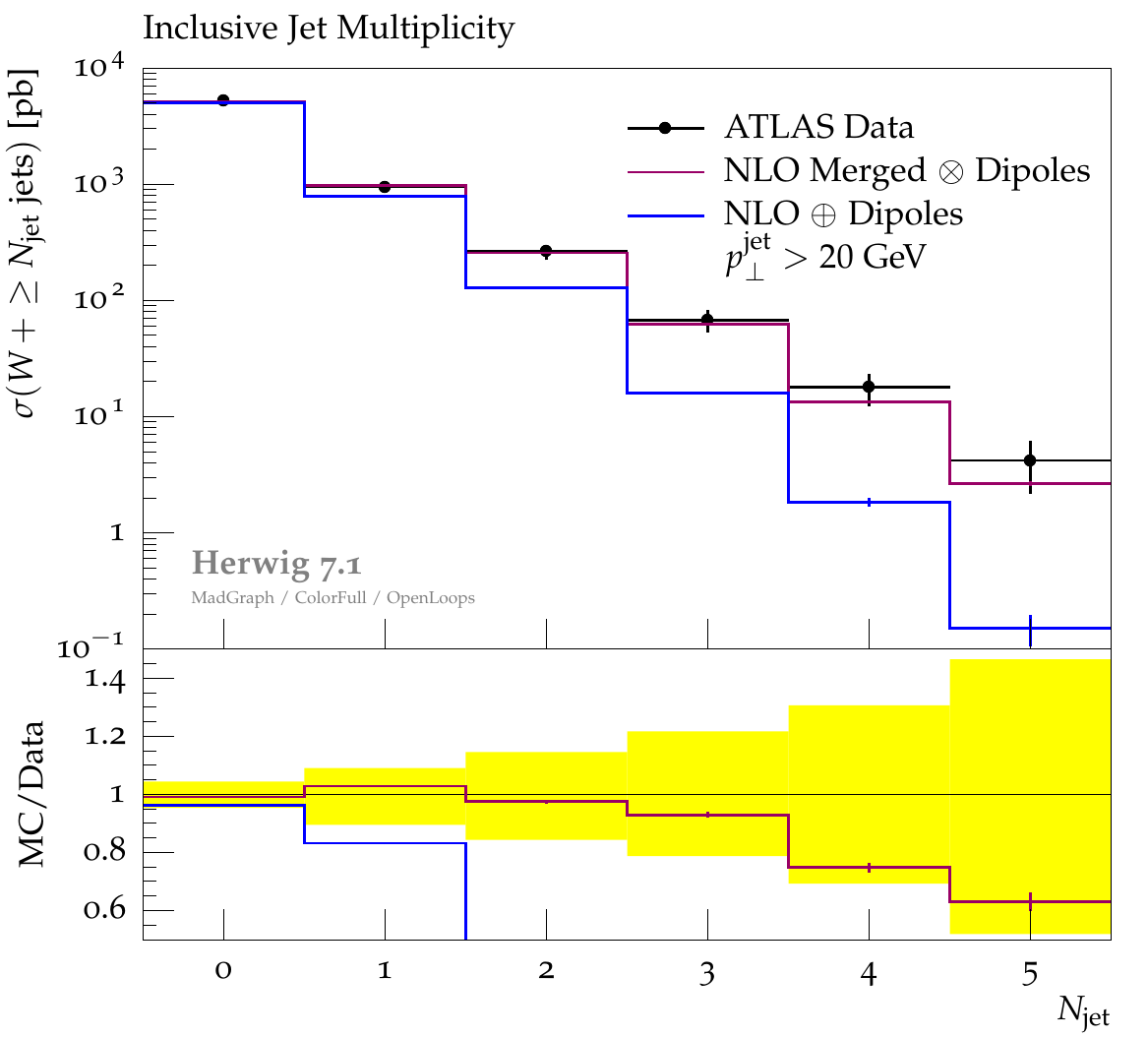}}
  \caption{The normalized $Z$ $p_\perp$ spectrum (top panel), and jet
    multiplicites in $W$ plus jets events (lower panel) as measured by ATLAS
    \cite{Aad:2011gj,Aad:2012en} and comparing the NLO matched prediction with
    the dipole shower to the NLO multijet merged prediction. Higher jet
    multiplicities and less inclusive quantities will receive bigger
    corrections through the merging algorithm. For these results we
    have used our run-time interfaces to \mbox{\textsf{MadGraph5\_aMCatNLO}}
    \cite{Alwall:2014hca} and \textsf{OpenLoops} \cite{Cascioli:2011va} to
    evaluate scattering amplitudes for each phase space point, and
    \textsf{ColorFull} \cite{Sjodahl:2014opa} to perform the colour
    algebra.}
  \label{fig:merging}
\end{figure}%  

Compared to the simple input file structure of the
\matchbox\ framework, minor additional commands are needed to perform
calculations with several jet multiplicities merged to the dipole
shower. Input file examples for a range of processes are provided in
\texttt{share/Herwig/Merging}.  Different from the standard NLO matching input
files for use with \matchbox, merging only requires a slightly different
process definition. For example,
\begin{quote}\tt \small
do MergingFactory:Process p p -> W+ [j j j]\\
set MergingFactory:NLOProcesses 2\\
set Merger:MergingScale 10.*GeV
\end{quote}
sets up
on-shell $W^+$ production with up to three jets and including NLO QCD
corrections to the inclusive and one-jet process. For the merging scale we
recommend some default ranges (LHC at $13\ {\rm TeV}$: $10 - 30\ {\rm GeV}$,
LEP at $91\ {\rm GeV}$: $4 - 6\ {\rm GeV}$ and for HERA run 2 with $27\ {\rm
  GeV}$ electrons/positrons on $820\ {\rm GeV}$ protons we have found that a
merging scale between $8 - 15\ {\rm GeV}$ has provided reliable
results). For colliders running significantly outside
these parameters, and in dependence on acceptance cuts, the value needs to be
adjusted, possibly down to small merging scales. This provides stable
predictions due to the unitarisation procedure.

Example plots are shown in Fig. \ref{fig:merging}, highlighting the fact that
inclusive quantities do not receive big corrections, while higher jet
multiplicities are significantly improved by the procedure. Variations of the
factorization and renormalization scales can be obtained as with all other
simulation setups.
  
\section{New Soft Model}  

Our model of soft interactions in the context of multiple partonic
interactions (MPI) has been replaced by a new approach. The existing MPI
model still forms the basis of the physics simulation by separating hard and
soft interactions with the help of the parameter $p_\perp^{\rm min, 0}$
\cite{Bahr:2008dy,Bahr:2008wk,Bahr:2009ek}.  In the context of this framework
a number of soft interactions $N_{\rm soft}$ is determined as before. The
model for colour reconnections \cite{Gieseke:CRmodel} stays in place and still
has a significant impact on the final state as previously uncorrelated
multiple scatters have to acquire some colour correlation.
  
\begin{figure}[t]
  \centering
  \scalebox{.7}{\includegraphics{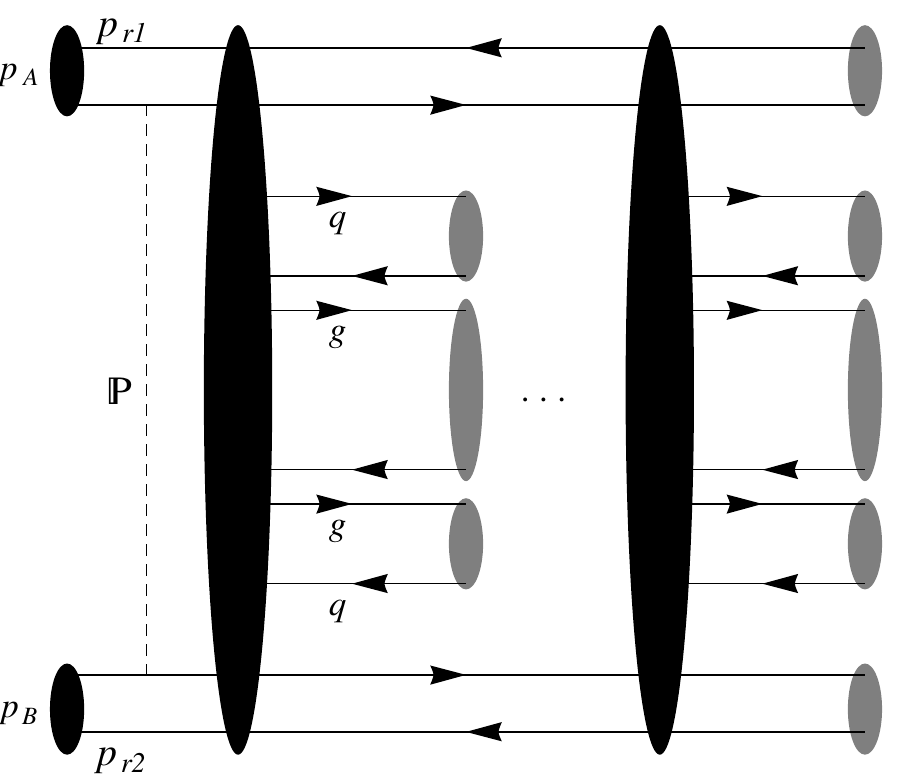}}
  \caption{  \label{fig:softmodel1}
    Colour structure of soft particles produced with the new
    model for soft interactions, shown in the context of a complete
    hadron-hadron interaction.  }
\end{figure}%

Instead of the generation of a gluon pair for each soft interaction we
now generate a number of rapidity-ordered gluons (and a pair of quarks)
as depicted in Fig.~\ref{fig:softmodel1}.  The gluons' rapidities are
determined based on multiperipheral kinematics \cite{Baker:1976cv},
where the longitudinal momenta are slightly smeared,
cf.~\cite{Gieseke:2016fpz}.  The unordered transverse momenta $p_\perp$
of the gluons are limited to the regime of soft transverse momenta,
i.e.\ $p_\perp < p_\perp^{\rm min, 0}$. Further details of the model are
described in \cite{Gieseke:2016fpz}.

\begin{figure}[t]
  \centering
  \scalebox{.7}{\includegraphics{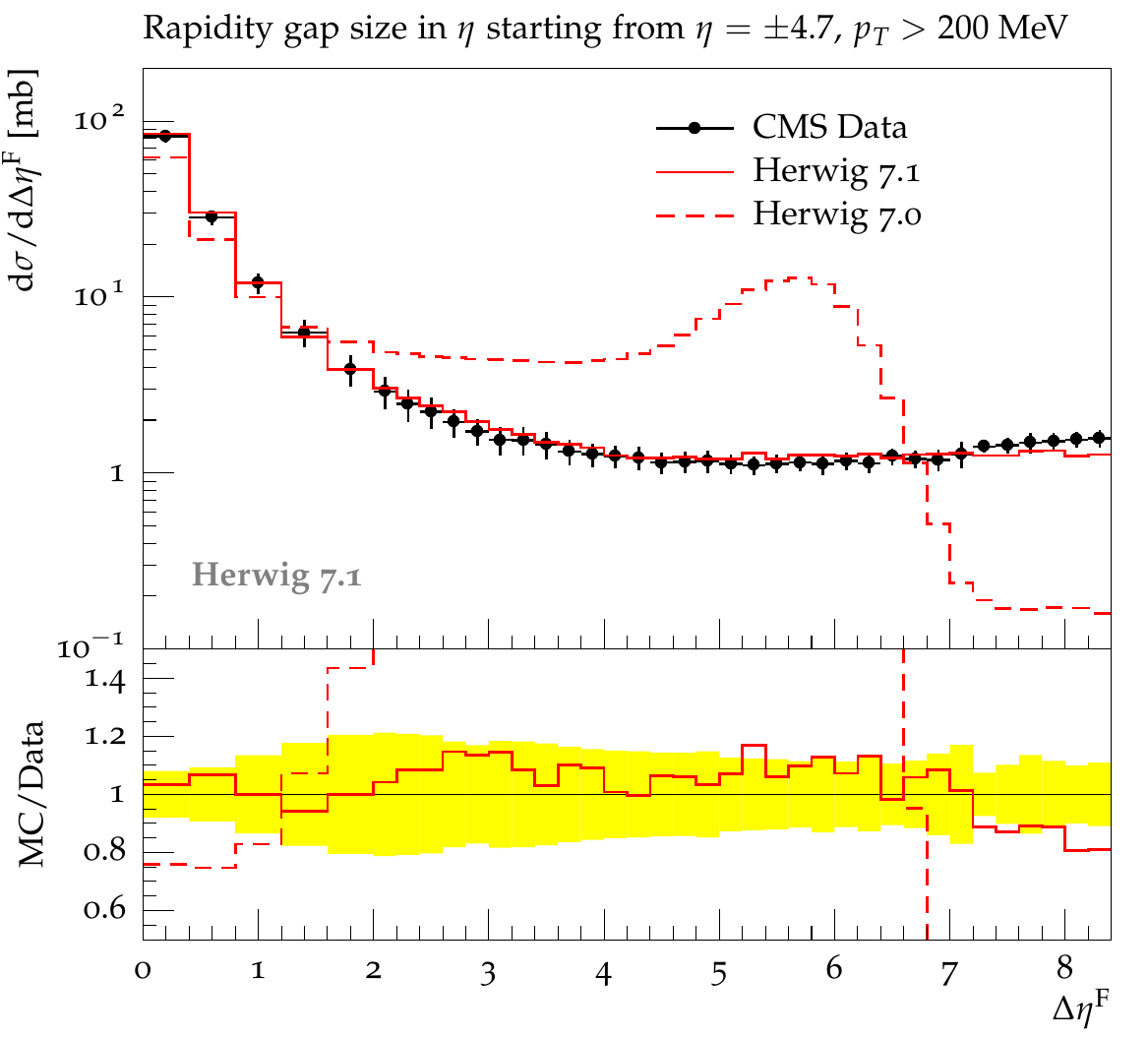}}
  \caption{The distribution of forward rapidity gaps with the new model
    including a model for diffractive final states (Herwig 7.1), compared to
    the old model (Herwig 7.0) and CMS data
    \cite{Khachatryan:2015gka}. \label{fig:softmodel2}}
\end{figure}

\begin{figure}[t]
  \centering
  \scalebox{.7}{\includegraphics{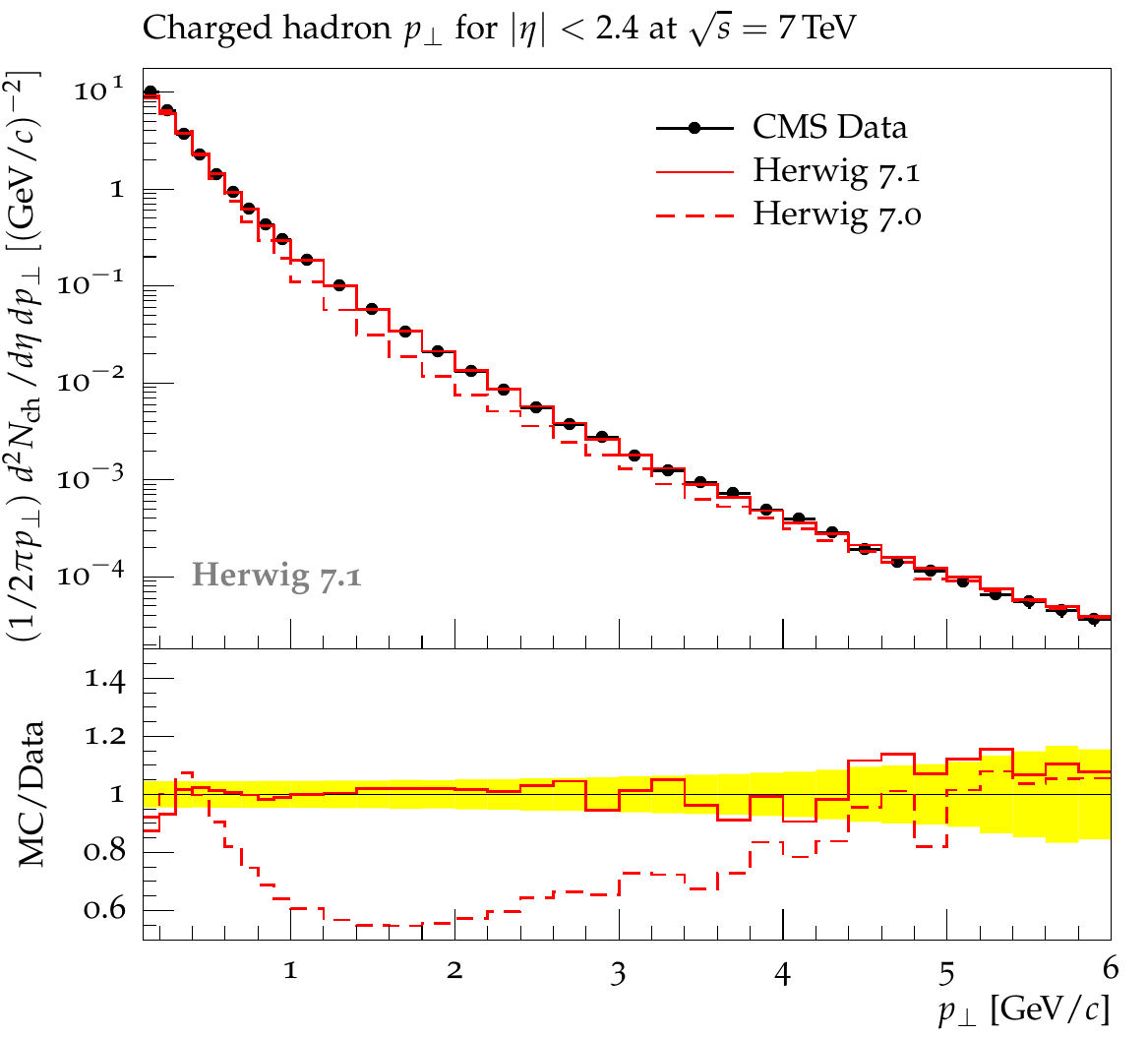}}
  \caption{Low transverse momentum spectrum of charged particles in
    non-single diffractive events with our old and new models for soft
    interactions compared to CMS data
    \cite{Khachatryan:2010us}.  \label{fig:softmodel3} }
\end{figure}

In order to complete the model towards softer and more forward interactions we
also added a simple model for diffractive scattering which complements the
hard MPI model for minimum-bias interactions.  The model for diffractive final
states heavily uses the cluster hadronization model already used by Herwig.
Details of the model and several results have been presented elsewhere
\cite{Gieseke:2016fpz,Gieseke:2017yfk,Gieseke:2017ccw}.  Here, we highlight
two findings: most notably, the unphysical predictions for the distribution of
forward rapidity gaps is now replaced by an excellent description of data,
Fig.~\ref{fig:softmodel2}, highlighting the expected composition of
non-diffractive events at small gap sizes, and diffractive contributions at
large gaps. We stress the fact that the old model, which generated a 'bump'
structure in this spectrum due to artificial colour re-connections, was not
meant to describe these interactions, so no
conclusions from this data comparison could be drawn. Another positive result
of the new modelling of soft particle production is the improvement of soft
transverse momentum spectra of charged particles, also in minimum-bias
interactions, see Fig.~\ref{fig:softmodel3}.

The model for soft interactions has become the new default model.  The
matrix elements for diffractive scattering are used alongside the hard
and soft MPI model by default in the simulation of minimum-bias matrix
elements.  There are two new parameters for the soft interaction model
that determine the number of gluons per soft interaction and its growth
with energy.  Both parameters have been tuned to minimum-bias data.

\section{EvtGen interface} 

\begin{figure}[t]
  \centering
  \scalebox{.7}{\includegraphics{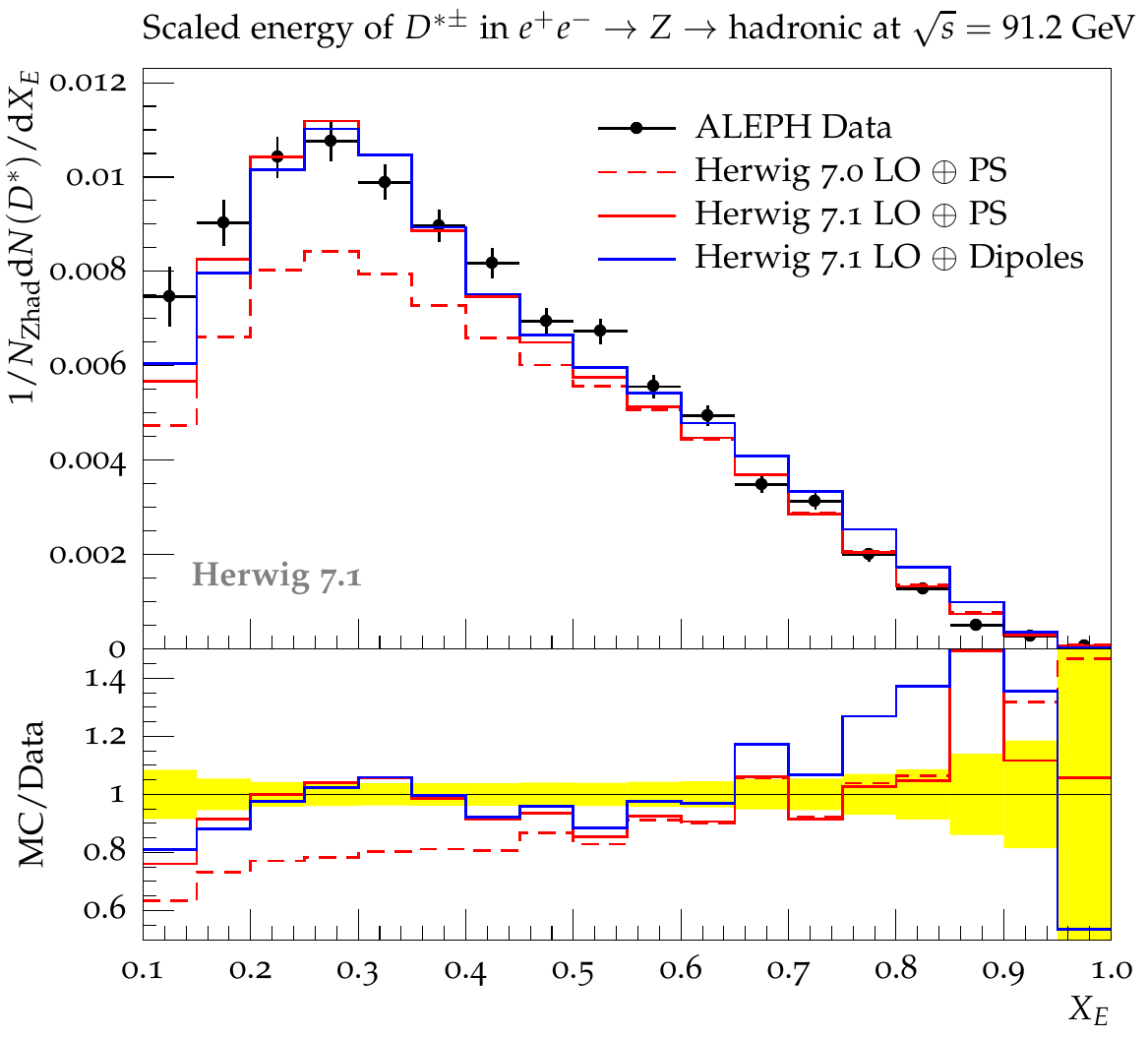}}
  \caption{The spectrum of $D^*$ mesons measured by the ALEPH experiment
    \cite{Barate:1999bg} compared to \Hw. As an example, we show LO plus PS
    predictions, however as expected these are not significantly changed in
    the presence of higher order corrections.}
  \label{fig:charm}
\end{figure}

The internal \Hw\ modelling of hadron decays includes sophisticated modelling
with off-shell effects and spin correlations. However, it has proven
impossible to provide a good tune to data for the decay of bottom and charm
mesons, largely due to the lack of published distributions.  Given that
\evt\cite{Lange:2001uf} has been tuned to non-public data from the B-factory
experiments and internally uses similar algorithms to include spin correlations
in particle decays, in \Hw\,7.1 we include an interface to \evt\ which
communicates the spin information between the two programs ensuring that the
full correlations are generated. \evt\ is now the default for the decay of
bottom and charm mesons. As there is less data for bottom and charm baryons
and our modelling of baryonic form-factors is more sophisticated, the decays of
heavy baryons continue to be performed by \Hw. This leads to the improvement
of a number of distributions, {\it e.g.} the momentum distribution of $D^*$
mesons~\cite{Barate:1999bg}, Fig.\,\ref{fig:charm}, where there is a
significant contribution from $D^*$ mesons produced in bottom meson decays.

\section{Mass Effects in the Dipole Shower}

\begin{figure}[t]
  \centering
  \scalebox{.7}{\includegraphics{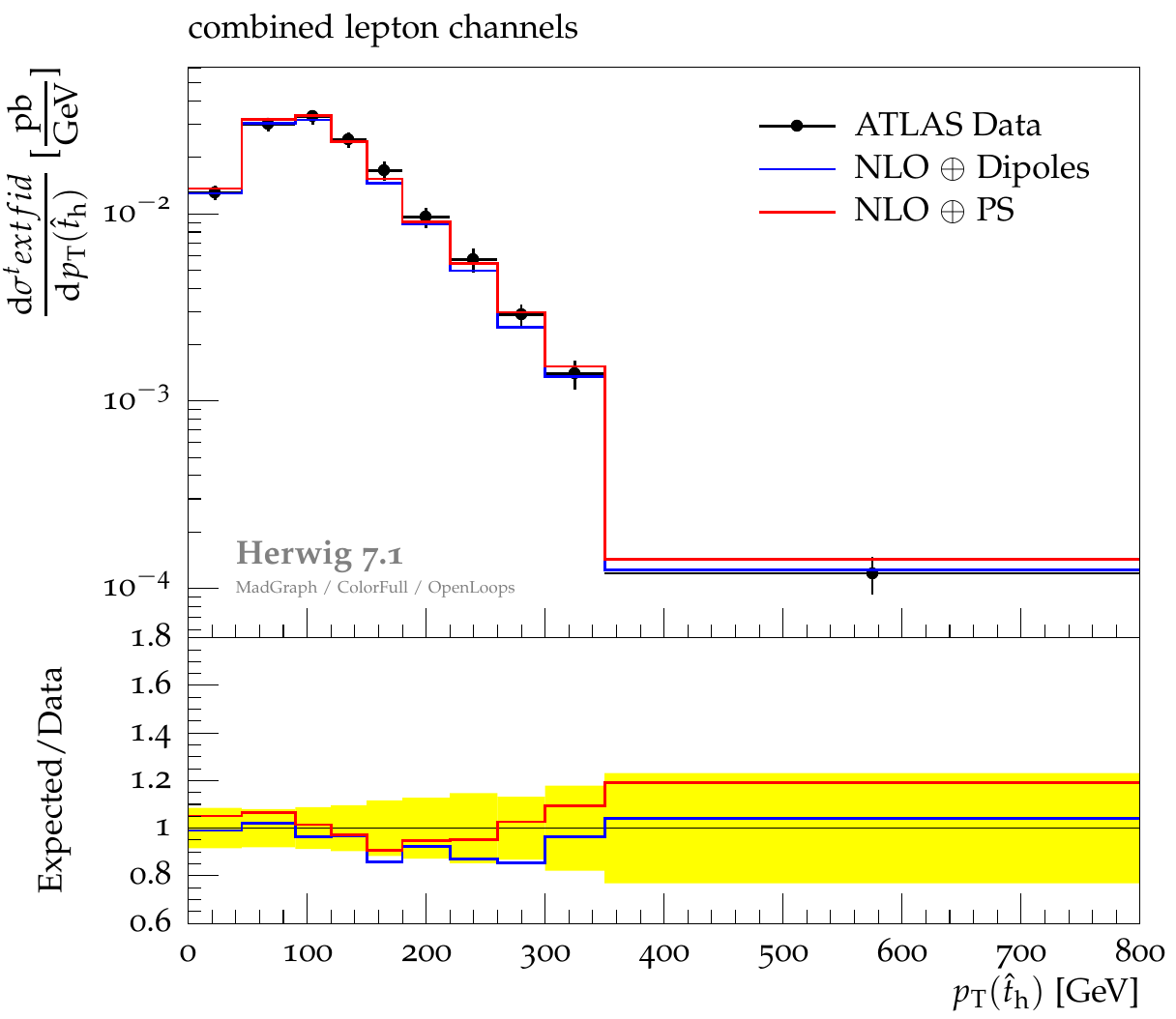}}
  \scalebox{.7}{\includegraphics{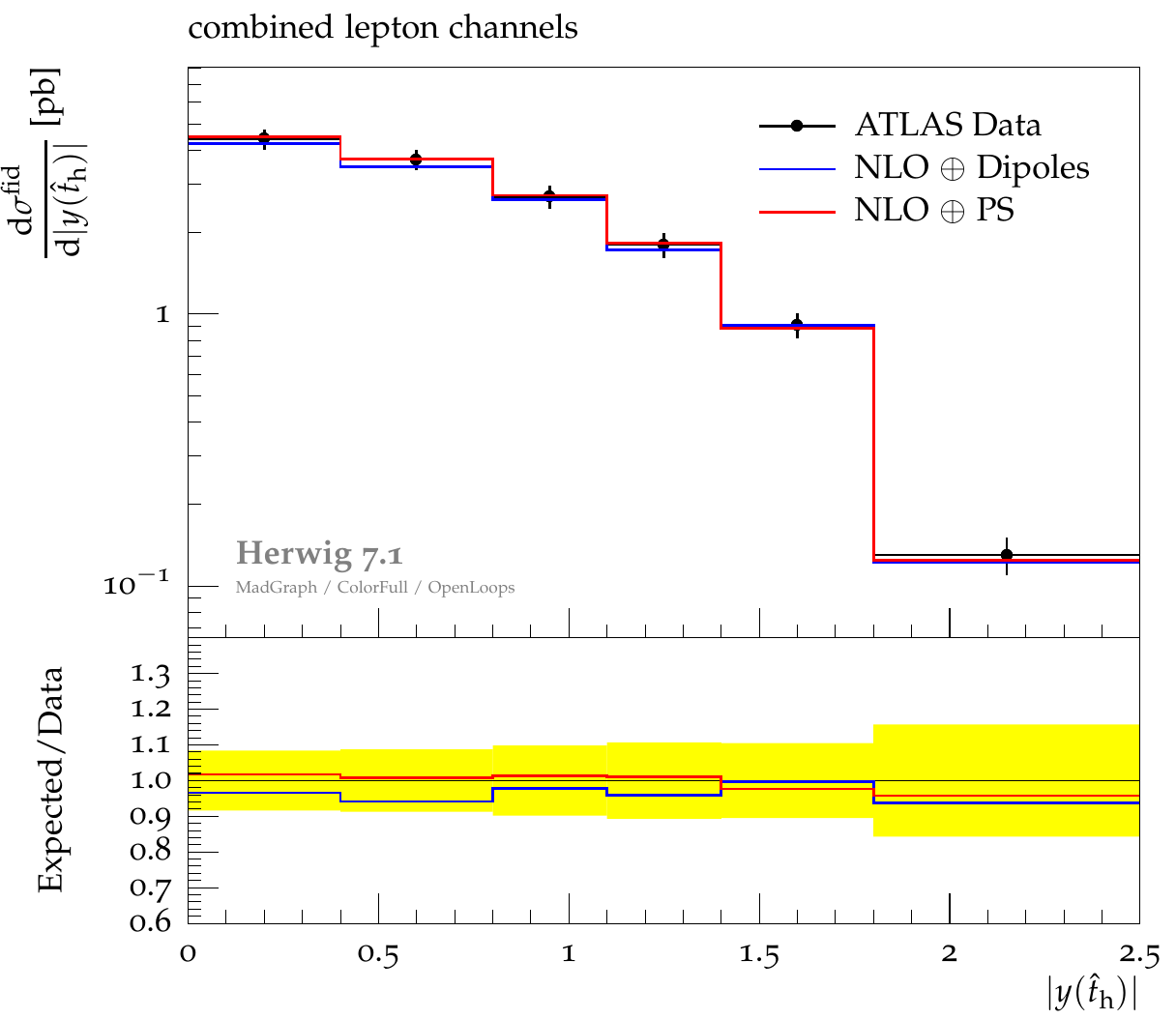}}
  \caption{Properties of top pair production in 7~TeV collisions at the LHC, as
    measured by ATLAS\cite{Aad:2015eia} and predicted by the
    QTilde and Dipole showers using the NLO+PS setup of \matchbox\ in
    \Hw\ 7.1.  More details will be presented in a forthcoming publication
    \cite{Cormier:2017xxx}. For these results we have used our
    run-time interfaces to \mbox{\textsf{MadGraph5\_aMCatNLO}} \cite{Alwall:2014hca}
    and \textsf{OpenLoops} \cite{Cascioli:2011va} to evaluate scattering
    amplitudes for each phase space point, and \textsf{ColorFull}
    \cite{Sjodahl:2014opa} to perform the colour algebra.}
  \label{fig:tprod}
\end{figure}

We recall that \Hw\ 7 contains two shower algorithms, based on angular ordering
(which we call QTilde) or dipole showering respectively. The dipole shower has
been extended in version~7.1 to include the showering of top
quarks in both their production and decay with the option to include the NLO
correction to the decay.
We show an example of the results for top production in Fig.~\ref{fig:tprod},
in comparison with ATLAS data\cite{Aad:2015eia}.
The dipole shower can now perform showering of all
Standard Model (SM) processes, including the NLO Powheg-type correction to all
SM decays. The NLO correction can be switched on and off by setting,
\begin{quote}\tt \small
	set DipoleShowerHandler:PowhegDecayEmission Yes/No
\end{quote}
and is on by default.

\begin{figure}[t]
  \centering
  \scalebox{.7}{\includegraphics{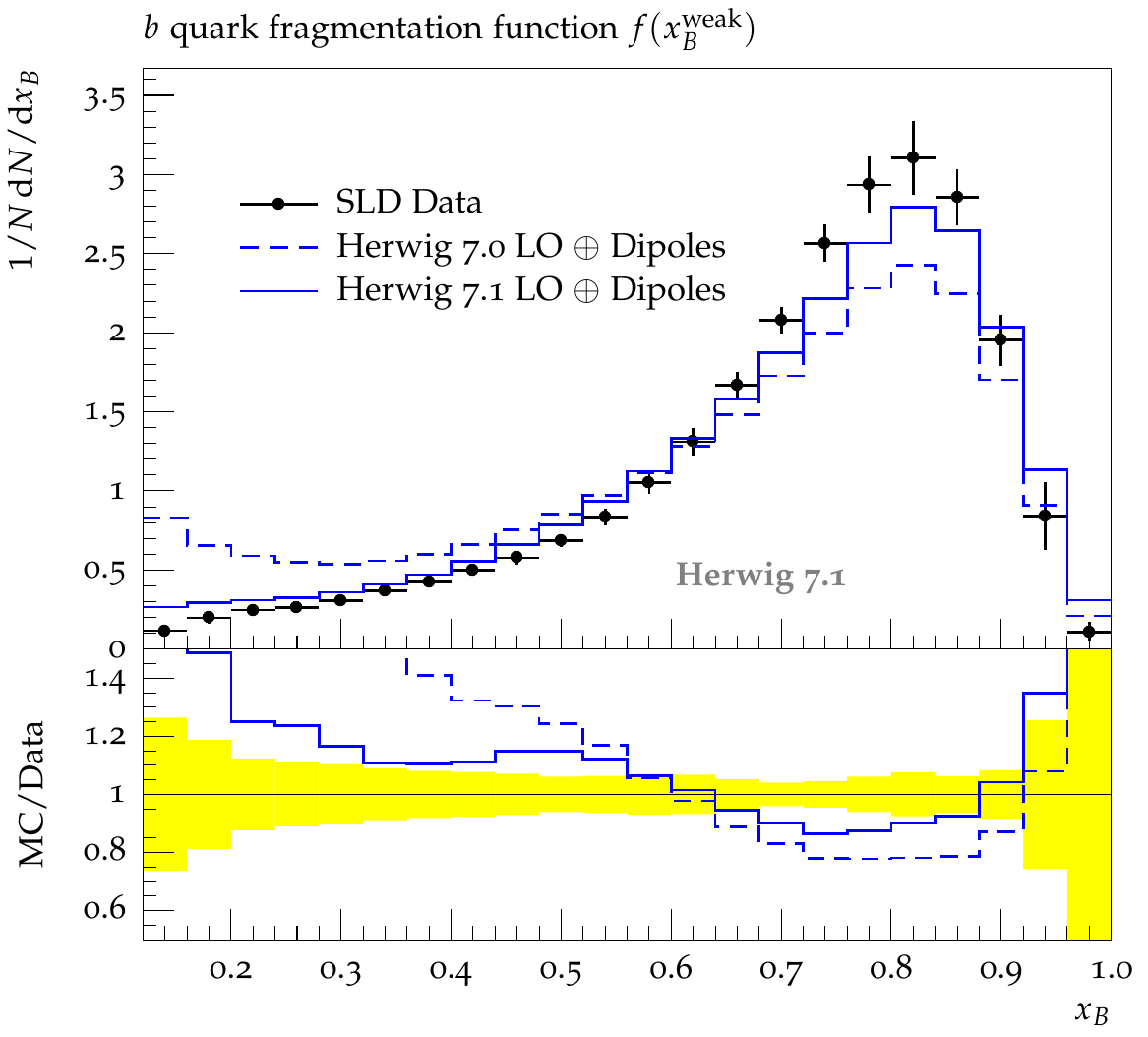}}
  \caption{The $B$-fragmentation as measured by SLD\cite{Abe:2002iq}
    and predicted by the dipole shower with the
    improved kinematics for massive quarks. More details will be presented in
    a forthcoming publication \cite{Cormier:2017xxx}.}
  \label{fig:bfrag}
\end{figure}

We have also performed a detailed analysis and new derivation of the
kinematics used to describe splittings of dipoles involving massive emitters
and/or spectators. As part of this we have derived and implemented covariant
formulations of the physical momenta of the partons following a splitting in
terms of the physical momenta of the partons prior to the splitting for these
dipoles. The kinematics for all dipole splittings in the dipole shower and
\matchbox\ now use such a formulation, with an evolution variable which is
directly connected to the transverse momentum variable relevant for the
collinear or quasi-collinear limits. The effect of these improvements can be
clearly seen in our modelling of $B$-Fragmentation in $e^+e^-$ annihilation at
the $Z^0$~mass, see
Fig.~\ref{fig:bfrag}, and more details will be covered in a forthcoming
publication.

\section{Shower Variations and Reweighting}
\label{sections:showervariations}

Evaluation of shower uncertainties is an important part of modern Monte Carlo
studies. Shower uncertainties are traditionally evaluated by performing a full
set of event simulations for each variation of interest.

To reduce the computational cost of evaluating shower uncertainties we have
introduced functionality to perform on-the-fly parton shower reweighting in
\Hw\ \cite{Bellm:2016voq}.  In this framework, each event is showered using a
central set of parameters. In addition, on a splitting-by-splitting basis, we
evolve a weight relative to the central shower for each set of varied
parameters. We have currently implemented reweighting to evaluate variations
of the factorization and renormalization scales used in the shower however it
is a general technique that could be applied to other variations in future
developments.

A very efficient sequence of the veto algorithm for the central scale choice
can lead to inefficient performance of the algorithm for the variations. We
have included a `detuning parameter' which can be used to improve the
convergence of the reweighted results at the expense of a less efficient
algorithm for the central prediction.

Reweighting is available in both showers. Multiple variations can be included
in a single run and each variation requires a unique name, `varName', which is
used to identify the weight in the \HepMC\ record. Each variation corresponds
to a pair of scale factors, $\xi_R$ and $\xi_F$, to be applied to the
renormalization and factorization scales respectively. Finally each variation
can be applied to the showering of the hard process only (Hard), secondary
processes only (Secondary) or to both parts (All):
\begin{quote}\tt \small
	do ShowerHandler:AddVariation VarName xR xF {Hard/Secondary/All}
	\\
	set SplittingGenerator:Detuning Factor
\end{quote}

\begin{quote}\tt \small
	do DipoleShowerHandler:AddVariation VarName xiR xiF {Hard,Secondary,All}
	\\
	set DipoleShowerHandler:Detuning Factor
\end{quote}

On top of using reweighting for the shower variations, the dipole shower
offers a number of reweighting and biasing facilities which are {\it e.g.}
used for the \KrkNLO\ method (see below). These are available through the
\texttt{DipoleSplittingReweight} and \texttt{DipoleEventReweight}
classes. Very flexible veto functionality is also available for the angular
ordered shower through the \texttt{ShowerVeto} and \texttt{FullShowerVeto}
classes.

\section{KrkNLO}

This version of \Hw\ contains an implementation of the
\KrkNLO\ method~\cite{Jadach:2015mza}.  This provides NLO QCD corrections to
LO matrix elements for specific processes following this paradigm as an
alternative to the other matching schemes available. The implementation
currently supports the Drell-Yan ($Z/\gamma^*$) process, and Higgs production
via gluon-fusion (in the large top-mass limit) and is available for the dipole
shower~\cite{Platzer:2009jq,Platzer:2011bc}. For the Drell-Yan process, it is
possible to use both the MC and ${\rm MC}_{\rm DY}$ variants of the MC
scheme\cite{Jadach:2016acv}. This module was validated against a previous,
independent, implementation using the published DY results of
Ref.~\cite{Jadach:2015mza} and was also used to simulate the first results for
this method in Higgs production \cite{Jadach:2016qti}. \KrkNLO\ can
be enabled by using
\begin{quote}\tt \small
  read Matchbox/KrkNLO-DipoleShower.in\\
  set KrkNLOEventReweight:Mode H\\
  set KrkNLOEventReweight:PDF MC\\
  set KrkNLOEventReweight:AlphaS\_R Q2\\
  set KrkNLOEventReweight:AlphaS\_V M2\\
\end{quote}
in combination with an MC-scheme PDF. The MC-scheme PDFs,
example input-cards, and other relevant codes are hosted at
\texttt{\href{https://krknlo.hepforge.org/}{https://krknlo.hepforge.org/}}.

\section{Other Changes}

Besides the major physics improvements highlighted in the previous sections,
we have also made a number of smaller changes to the code and build system
which we will summarize below. Please refer to the online documentation for a
fully detailed description or contact the authors.

\subsection{Steering, input files and weights}

The steering of the \Hw\ executable has seen a number of improvements, mainly:
\begin{itemize}
\item A new \texttt{run} mode has been added to solely perform the merging of
  integration grids from parallel integration runs,
\begin{quote}\tt \small
\vspace*{1ex}
Herwig mergegrids <run file name>
\vspace*{1ex}
\end{quote}
\item A high-level run-time interface is now available to steer \Hw\ within
  more complex frameworks such as experimental software without the need to
  execute the binary. This includes all of the \texttt{read}, \texttt{build},
  \texttt{integrate}, \texttt{mergegrids} and \texttt{run} steps.
\end{itemize}
The structure of input files for non-\matchbox-based processes has
been adapted to use the {\it snippet} input file mechanism and is now in line
with steering matched and merged processes. On top of this, a large number of
input file switches which have before used \texttt{On,Off} or
\texttt{True,False} to indicate their state have been changed to
\texttt{Yes,No}.

As far as integration and event generation are concerned, we have made a
choice that by default sampling is run in \texttt{AlmostUnweighted} mode, {\it
  i.e.} events carry in general varying weights, most of which are unity.
This is to account for the fact that the grid adaption might only
have encountered a maximum weight close to the true maximum weight and strict
unweighting in this case could skew distributions and cross section
estimates. The reference weight to which events are unweighted can also be
adjusted to keep weight distributions mostly narrow while reducing
fluctuations in tails due to a small frequency of contributing
events. Alongside this, the adaption parameters of both the
\texttt{CellGridSampler} and \texttt{MonacoSampler} have been revised.

\subsection{Minor improvements and bug fixes}

A number of minor changes and bug fixes are worth noting, in particular, there
have been new options for the physics simulation besides the ones described in
the previous text:
\begin{itemize}
\item Colour reconnection of octet systems into a single cluster are now
  prevented, improving the description of a number of observables sensitive to
  these dynamics, as well as some unexpected features which have been observed
  in preceeding work \cite{Gras:2017jty}.
\item For both showers it is now possible to alter the scale choice and
  ordering properties in $g\to q\bar{q}$ splittings.
\item New options of shower scale choices are available for NLO matched
  processes.
\item The $\alpha_s$ running in the dipole shower can now explicitly be
  switched to use the CMW scheme \cite{Catani:1990rr}, through both a scaling
  of its argument as well as by explicitly adding the $\alpha_s^2 K_g$
  contribution, such that these contributions do not anymore need to be
  absorbed into a tuned value of $\alpha_s$.
\item Several options have been added for the emission phase space in the
  dipole shower, which are subject to a more detailed, future study.
\item Structures in \ThePEG\ have been extended to cover processes which do
  not exhibit a (tree) diagram-like internal structure, such as instanton- and
  sphaleron-induced transitions.
\end{itemize}
Technical issues which have been addressed include:
\begin{itemize}
\item \matchbox\ is now able to handle processes which do not contain coloured
  external legs.
\item The dipole shower can handle zero-momentum-transfer initial-final colour
  connections, which have prevented running minimum bias simulation with this
  shower algorithm before.
\item Several levels of assumptions (such as Standard Model-like interactions,
  conservation of lepton flavour number, quark flavour diagonal interactions)
  can be imposed on the generation of candidate sub-processes to reduce
  combinatorial complexity for processes with many legs.
\end{itemize}

\subsection{Build and external dependencies}

As of version 7.1, \Hw\ is now enforcing the use of a C++11 compliant
compiler, and C++11 syntax and standard library functionality is used widely
within the code. The \texttt{herwig-bootstrap} script is able to provide such
a compiler along with a full \Hw\ plus dependencies build.
\texttt{herwig-bootstrap} will also enforce the newest versions of external
amplitude providers; specifically we now use:
\begin{itemize}
\item \OL\ \cite{Cascioli:2011va} versions $\ge$ 1.3.0 with the
  \Collier\ library \cite{Denner:2016kdg} for tensor reduction (should older
  versions of \OL\ be required, the input files require the additional option
  \texttt{set OpenLoops:UseCollier Off}), and
\item \GoSam\ versions $\ge$ 2.0.4 to pick up the correct normalization for
  loop induced processes outside of specialized setups.
\end{itemize}
A number of changes have also been implemented to reduce run-time load for
allocating and de-allocating various containers, and to reduce overall memory
consumption.

\subsection{Licensing}

While \Hw\ 7.0 and older versions of \HWPP\ have been licensed under the GNU
General Public License GPL, version 2, \Hw\ 7.1 and future versions are
distributed with the GPL version 3. The MCnet guidelines for the distribution
and usage of event generator software in an academic setting apply as before,
and both the legally binding GPL license and the MCnet guidelines are
distributed with the code.

\section{Summary and Outlook}

We have described a new release, version 7.1, of the \Hw\ event
generator. This new release contains a number of improvements to both
perturbative and non-perturbative simulation of collider physics and will form
the basis of further improvements to both physics and technical aspects.

\section*{Acknowledgments}

We are indebted to Leif L\"onnblad for his authorship of \ThePEG, on which
\Hw\ is built, and his close collaboration, and also to the authors of
\Rivet\ and \textsf{Professor}. We are also grateful to Malin Sj\"odahl for
providing the \textsf{ColorFull} library for distribution along with \Hw.

This work was supported in part by the European Union as part of the FP7 and
H2020 Marie Sk\l odowska-Curie Initial Training Networks MCnetITN and MCnetITN3
(PITN-GA-2012-315877 and 722104), the Lancaster-Manchester-Sheffield Consortium
for Fundamental Physics under STFC grant ST/L000520/1 and the Institute for
Particle Physics Phenomenology under STFC grant ST/G000905/1. AP acknowledges
support by the ERC grant ERC-STG-2015-677323.
MHS acknowledges support of this project by the Institute for Particle Physics
Phenomenology through an IPPP Associateship.
SP is grateful to the Erwin-Schr\"odinger-Institute at Vienna for kind
hospitality while part of this work was completed. 

CR acknowledges support by the US Department of Energy under grant DE-SC0010102.

\bibliography{Herwig}
\end{document}